# Comparative study of different scattering geometries for the proposed Indian X-ray polarization measurement experiment using Geant4

S. V. Vadawale[a], B, Paul[b], J. Pendharkar[a], Sachindra Naik[a]

**a:** Physical Research Laboratory, Navarangpura, Ahmedabad, India – 380 009
**b:** Raman Research Institute, Sadashivnagar, Bangalore, India – 560 080

**Abstract**

Polarization is a very important property of radiation from astrophysical sources. It carries unique information regarding the emission mechanism, physical conditions as well as emission geometry at the origin. Polarization measurements in X-rays can provide unique opportunity to study the behavior of matter and radiation under extreme magnetic fields and extreme gravitational fields. Unfortunately, over past two decades, when X-ray astronomy witnessed multiple order of magnitude improvement in temporal, spatial and spectral sensitivities, there is no (or very little) progress in the field of polarization measurements of astrophysical X-rays. Recently, a proposal has been submitted to Indian Space Research Organization (ISRO) for a dedicated small satellite based experiment to carry out X-ray polarization measurement, which aims to provide the first X-ray polarization measurements since 1976. This experiment will be based on the well known principle of polarization measurement by Thomson scattering and employs the baseline design of a central low Z scatterer (typically Lithium, Lithium Hydride or Beryllium) surrounded by X-ray detectors to measure the angular intensity distribution of the scattered X-rays. The sensitivity of such experiment is determined by the collecting area, scattering and detection efficiency, X-ray detector background, and the modulation factor. Therefore, it is necessary to carefully select the scattering geometry which can provide the highest modulation factor and thus highest sensitivity within the specified experimental constraints. The effective way to determine optimum scattering geometry is by studying various possible scattering geometries by means of Monte Carlo simulations. Here we present results of our detailed comparative study based on Geant4 simulations of five different scattering geometries which can be considered within the weight and size constraints of the proposed small satellite based X-ray polarization measurement experiment.

## 1. Introduction:

Since the birth of X-ray astronomy in early 1960s, there has been tremendous improvement in the sensitivity of the X-ray astronomical observations. As a result, three of the four dimensions of X-ray astronomy viz. photometry, imaging and spectroscopy are very well developed and fully mature subjects. However, the fourth dimension, namely polarimetry, so far has been almost untouched observationally. There has been only one measurement of polarization in X-ray astronomy, which was carried out for Crab nebula about 30 years ago. The OSO-8 satellite carrying an X-ray polarimeter was launched by NASA in 1976 which measured ~19% polarization at 2.6 keV and 5.2 keV for the Crab nebula [1]. The only other polarimeter ever launched onboard satellite Arial-5 and few earlier rocket and balloon-borne experiments did not result in successful measurement [2,3,4,5]. The main reason for the lack of X-ray polarization measurements so far is their extremely photon hungry nature that severely limits the sensitivity of the instruments. This coupled with the limitations of the measurement techniques resulted in almost three decade void in X-ray polarization measurements.

The X-ray polarization measurements are very important because they provide two independent parameters i.e. degree and angle of polarization to constrain the physical model for the X-ray source. Such measurements can provide unique opportunity to study the behavior of matter and radiation under extreme magnetic and gravitational fields. Prime targets for X-ray polarimetric observations are neutron stars

(isolated neutron stars as well as neutron stars in binary systems) where the X-ray polarization measurements can provide direct information of the intensity and geometry of the magnetic field. Polarimetric observations of accreting Galactic Black Hole systems are also very interesting because they provide unique opportunity to test some of the predictions of general relativity which are inaccessible by any other means. Another set of interesting targets for polarization observations are the cosmic acceleration sites such as supernovae remnants and jets in Active Galactic Nuclei (AGN) or micro-quasars. Polarization observations of these sites will provide direct information about the geometry of the shocked sites as well as the structure and intensity of magnetic fields therein.

The importance of X-ray polarimetric observations was realized for long time. There have been many reports on theoretical prediction of X-ray polarization from various types of X-ray sources [6] even during early years of X-ray astronomy [7]. Particularly in recent times, requirement of X-ray polarimetric observations has been strongly realized, mainly because, in various classes of X-ray sources, even the best spectroscopic or photometric observations can not remove the degeneracy in the theoretical models [8]. As a result, several groups world wide have been attempting experiments which can provide the meaningful X-ray polarimetric observations. With the recent technological advances, it has now become possible to carry out such observations. Particularly the advances in the photo-electron tracking technique has lead to the real probability of having an X-ray polarimeter as one of focal plane instrument for next generation X-ray observatories[9,10]. The present state of X-ray polarimetry as an emerging field is well described in various articles in literature [11,12].

Recently we have proposed an exploratory X-ray polarization measurement experiment to the Indian Space Research Organization (ISRO) [13]. This will be based on the well known principle of polarization measurement by Thomson scattering and will be sensitive to some of the bright X-ray sources. Here we present the results of our detailed Monte Carlo simulations, performed with Geant4 toolkit, in order to determine the optimum scattering geometry and to determine the sensitivity of our proposed experiment.

## 2. Proposed Experiment:

The X-ray polarization measurement experiment was proposed in response to the ISRO's announcement of opportunity for an experiment to be launched on a dedicated small satellite. The proposed X-ray polarimeter will be able to measure polarization in the 7 – 30 keV in most of the bright Galactic X-ray sources. The design goal is to achieve sensitivity of 2 – 3% Minimum Detectable Polarization (MDP) in a 50 – 100 mCrab bright X-ray source in an exposure of one million seconds. The experiment configuration will follow the standard Thomson scattering geometry of a central scatterer surrounded by X-ray detectors. We plan to use position sensitive proportional counters for measuring the azimuthal distribution of the scattered X-ray photons. The scatterer will be made up of a low Z element, Be or Li. Results of the present study as well as other practical consideration will lead to the final design of the scatterer as well as the proportional counter.

### 2.1 Rationale for choosing polarimeter design based on Thomson scattering:

There are three basic techniques for measuring linear polarization of the X-ray photons, namely Bragg reflection, Thomson scattering and photo-electron imaging. Bragg reflection is one of the oldest and clean techniques to measure polarization of X-rays. Both the X-ray polarimeters flown to space so far (onboard Arial-5 and OSO-8) were of this type. However, the major drawback of this technique is that it works only at discrete energies. This results in very low sensitivity. Photo-electron tracking is the latest technique for measuring polarization of X-rays. Very high resolution position sensitive X-ray detector required to image the photo-electron tracks (a few hundred microns in a gaseous medium) are becoming available only recently and have not been used so far in any space experiment. The main disadvantage of this technique, particularly in the context of the present ISRO announcement of opportunity is that they have very small collecting area. Therefore these detectors must be used with the X-ray focusing optics. This combination might provide the best sensitivity for X-ray polarization measurement. However, it requires a full fledge X-ray astronomy mission which is out of scope for the present opportunity. Therefore, we adopt a pragmatic approach to limit the sensitivity goal and use the well established technique of X-ray polarization measurement based on Thomson scattering.

## 2.2 Sensitivity of a Thomson scattering based X-ray polarimeter:

The basic geometry of an X-ray polarimeter based on Thomson scattering is shown in Figure 1. The incident X-ray photons are scattered from the central scatterer and the azimuthal distribution of the scattered photons is measured by the surrounding X-ray detectors. For polarized radiation, the differential Thomson scattering cross-section is given by following equation –

$$d\sigma = r_e^2 (1 - \sin^2\theta \cos^2\varphi) d\theta d\varphi$$

where θ is the angle between direction of the scattered photon and direction of the incident photon, φ is the angle between the scattering plane and the plane defined by the direction and electric field vector of the incident photon and $r_e$ is the classical electron radius. The resultant azimuthal distribution follows cos(2φ) pattern given as -

$$C(\varphi) = A\cos(2(\varphi - n + \frac{\pi}{2})) + B$$

Where φ and η are the azimuthal coordinate and polarization angle of the incident radiation respectively where as $A$ and $B$ are constants.
The modulation factor ($\mu$) for the pattern is expressed as

$$\mu = \frac{C_{max} - C_{min}}{C_{max} + C_{min}}$$

The degree of polarization for the incident radiation is defined as

$$P = \frac{\mu_P}{\mu_{100}}$$

Where $\mu_p$ is the observed modulation factor for the unknown polarization, p, and $\mu_{100}$ is the modulation factor of the instrument for the 100 % polarized radiation.
Sensitivity for the polarization measurement or the minimum detectable polarization (MDP) is defined as –

$$MDP = \frac{n_\sigma}{\mu_{100} A \varepsilon R_{src}} \sqrt{\frac{2(\varepsilon A_c R_{src} + A_d R_{bkg})}{T}}$$

Where $n_\sigma$ is the significance level (number of sigma), $R_{src}$ is the incident rate of photons per unit area, ε is the overall detection efficiency, $A_c$ is collecting area, $A_d$ is the total detector area, $R_{bkg}$ is the detector back ground count rate per unit area, $\mu_{100}$ is the modulation factor of the polarimeter for fully polarized radiation and $T$ is the integration time. Here all parameters except $\mu_{100}$ and ε can be determined independently. However, these two parameters must either be measured experimentally or must be determined by means of simulations.

## 2.3 Possible scattering geometries:

Above expression shows that in order to maximize the sensitivity (e.g. to achieve lowest MDP), it is necessary to maximize both the modulation factor as well as detection efficiency of the scattered photons. For higher scattering efficiency it is necessary to use scattering material of lowest possible Z, (because the cross-section of the competing photoelectric interaction is proportional to $Z^5$). Also, the depth of the scatterer in the direction of the incident photons must be sufficiently large to have significant probability for the primary Thomson scattering interaction. However, in order to obtain highest modulation factor, it is necessary to minimize multiple interactions within the scatterer volume itself. This means that the thickness of the scattering element in the direction of the incident photon must be large and its thickness in the perpendicular directions must be very small. However, such an ideal shape of a narrow tube for the scattering element is not practical because of its very small collecting area. Thus it is necessary to design the shape of the scattering element as well as the surrounding detector in order to optimize the collecting area, the modulation factor as well as efficiency of detection of the scattered photons.

Larger collecting area can be achieved by using a thick disc shaped scatterer, but in this case the modulation would be limited due to the fact that the photons scattered close to 90 degree has to pass through large distance within the scatterer itself. This can be avoided if the scatterer is kept at some angle e.g. a slanted plate, but this lacks azimuthal symmetry. The azimuthal symmetry can be restored by using multiple of such plates to make pyramid shape scatterer or even better with a conical scatterer. The X-ray detector surrounding the scatterer can have different possible geometries. For example, one possibility is to use only one (or two) detector to measure scattered photon in one direction and then the entire system can be rotated with respect to the viewing axis to cover all directions. The other possibility is to keep detectors in all directions surrounding the scatterer to simultaneously detect photons scattered in any azimuthal angle. This can be achieved by multiple flat detectors or a single cylindrical detector. These multiple possibilities for the experiment configuration have varying degree of complexity in terms of implementation. Therefore, in order to select appropriate scattering geometry, it is essential to have some idea of the limiting sensitivities in each case. We adopt the method of Monte Carlo simulations to carry out such comparative study between various possible scattering geometries to have an estimate of limiting sensitivities in each case. Though the final selection would be driven by many other practical considerations, these results would be helpful to make an informed decision.

We considered two options for the X-ray detector surrounding the scatterer – four square proportional counters and a single cylindrical proportional counter. For the scattering elements, we consider four different shapes viz. a hollow cone, hollow pyramid, circular disk and square slab. For cylindrical detector we use only circular scatterer e.g. disc and cone. Thus we have total six different scattering geometries. We carry out Monte Carlo simulation of Thomson scattering in each of these geometries to determine the modulation factor for 100 % polarized X-rays as well as their overall efficiency for detection of the scattered photons.

## 3. Geant4 Simulations:

Geant4 is a standard toolkit for simulation of interaction of radiation with matter involving highly complex geometries [14]. We have developed our application code using Geant4 toolkit (version 4.9.2) to simulate the interaction of the polarized X-ray photons and their detection in the detector material. All Geant4 applications require specifying the detector geometry and the physics processes. Individual particles are then randomly generated by a particle generator and tracked through the detector volumes. Typically Geant4 tracks the particle silently without producing any output; however, it provides necessary tools to extract almost all parameters of the particle tracking. The application has to extract the required parameters and record them. Details of the physics processes, geometry and tracking for our application are given in the following sections.

### 3.1 Physics Processes:

The Geant4 toolkit has wide range of physics processes (in some cases different implementation of the same process). Therefore, the application developer has to carefully select appropriate physics process relevant for the simulation. Since we are mainly concerned with the interaction of low energy polarized X-ray photons, we employ only the low-energy electromagnetic process. Specifically, we use *G4LowEnPolarizedPhotoElectric*, *G4LowEnPolarizedRayleigh*, *G4LowEnPolarizedCompton*, *G4LowEnBremss* and *G4LowEnIonization*. In order to verify that these processes, particularly the Rayleigh and Compton processes, yield appropriate azimuthal modulation for 100% polarized X-rays, we first carried out an experiment with the ideal scattering geometry with a very thin scatterer tube (as shown in Figure 1). In this case, all photons were incident at the center of the scatterer face and the azimuthal angle of the scattered photon. The results are shown in Figure 2. It can be seen that there is very good agreement between the simulated and expected modulation of the azimuthal distribution of scattered photon. We use the same physics list in all our later simulation.

### 3.2 Geometry construction:

This experiment is proposed for the small satellite platform of ISRO. Typical base area and payload weight allowed for this platform are ~60 × 60 $cm^2$ and ~100 kg, respectively. We have selected the dimensions for the simulated experiment configuration considering these constrains. We have constructed two types of proportional counters – square and cylindrical; four types of scatterer shape – hollow cone, hollow pyramid, disc, slab; and three different scattering materials - Beryllium (Be), Lithium (Li) and Lithium Hydride (LiH). In case of proportional counters, the effect of the input window (made of thin mylar as it is transparent at 5 keV and above) is ignored but the aluminum grid structure supporting the entrance window is considered in the simulation. The exact specifications of the two types of proportional counters as well as for the four types of scatter shapes are given in Table 1. The constructed geometries are shown in Figure 3. It should be noted that the values used for these simulations are not frozen for the final experiment configuration. The actual experiment dimensions can be different from those considered here. However, the basic idea behind the present simulations is to compare among different geometries for the same detector dimensions in case of square detectors and between square and cylindrical detector for similar collecting area. Thus it is expected that these results will hold true even if the overall detector dimensions change.

### 3.3 Particle Generation, Tracking and Recording:

We use General Particle Source (GPS), which is available in the standard Geant4 distribution as 5–30 keV polarized X-ray photon generator. The main advantage of using GPS is that almost all of the properties of the generated particles can be specified at run time rather than compile time which makes it easier to change the properties such as energy and polarization in between the runs. The particle generating surface was kept perpendicular to the instrument axis and photons were incident parallel to the axis. The shape of the particle generating surface was square and circular for the simulations with square and cylindrical detectors respectively. In case of simulations with square proportional counters, the actual collecting area changed with different scatterer shapes. However, the particle generator area was kept same across these in order to include the effect of the different collecting area on final sensitivity.

Geant4 tracks the particle through the experiment geometry in units of 'Steps'. The photons and secondary electrons are tracked until they deposit all their energy in any defined volume or go out of the 'World' volume which encompasses the entire experiment geometry. For every step, we store the physical process defining the step, physical volume of the step, energy deposition if any and secondary particle generation if any. These details are stored by means of "User Event Information" (class G4VUserEventInformation). At the end of every event, this information is analyzed to check whether the event had interaction in the proportional counters defined by active volume consisting of Xenon gas. For all events having interaction in the proportional counter, the complete event interaction history is written in an output file. This also includes details like number of scattering in the scatterer, electron escape from the volume, scattering in xenon medium etc. However, in our analysis we do not consider any other information for the event apart from the position of its interaction in the proportional counter.

### 3.4 Simulation Method:

For the present simulations, we consider two types of X-ray detectors namely, four square proportional counters (4PPC) and cylindrical proportional counter (CYLL), and four types of scatterer shapes namely, circular disc (DISC), square slab (SLAB), hollow cone (CONE) and hollow pyramid (PYRD). The physical dimensions of the proportional counters used in simulation are given in Table-1. The proportional counters are assumed to have Xenon gas filled at one atmosphere pressure. The slab and pyramid shapes of the scatterer are not applicable for the cylindrical proportional counter. Thus, we have total six different scattering geometries, which we refer to with acronyms – 4PPC-DISC, 4PPC-SLAB, 4PPC-CONE, 4PPC-PYRD, CYLL-DISC and CYLL-CONE. For each of these geometries, the thickness of the scatterer is not constrained and one of the objectives of these simulations is to determine the optimum thickness in different cases. Therefore, we carry out simulations for a range of thicknesses, 5-140 mm, for each of the scattering geometry. For each thickness, we run the simulation for a range of energies – 5 to 30 keV at a step of every 5 keV, and at each energy the polarization angle is varied in the range of 0 to 90 degree at a step of every 5 degree. For every combination of scattering geometry, scatterer thickness, energy and polarization angle, we carry out simulation for 1000000 photons incident on the scatterer and store the output for each photon detected in the proportional counter.

## 4. Data analysis:

The output of each simulation run is stored in the form of event list. Further analysis of this data is carried out separately by using IDL. Each event line contains the exact location of the interaction in the proportional counters. However, in real life the information of the interaction location will be limited by the number of cells in the proportional counters. The square proportional counters are taken to have eight anode wires. These detectors will be positioned with anode wires running parallel to the instrument axis. Thus total four proportional counters gives 32 bins for the azimuthal scattering angle. The cylindrical detector also gives 32 azimuthal bins. The first step in the data analysis is to determine the azimuthal scattering angle in the instrument reference frame for every event. This azimuthal angle also includes the effect of variation of the polarization angle of the incident photons Therefore, the next step is to determine the azimuthal scattering angle in common reference frame by correcting for the relative change in polarization angle (in real experiment also the relative polarization angle change due to rotation of the instrument). This is achieved by subtracting the polarization angle of the incident photons from azimuthal scattering angle of each event. After determining the corrected azimuthal scattering angle in a common reference frame, the 32 bin azimuthal distribution histogram is generated with appropriate azimuthal angle bins. The azimuthal distribution histograms for different polarization angles are then co-added to get the total azimuthal modulation pattern for a given scatterer thickness and photon energy. For the present analysis, we assume that no position information is available along the anode wire and thus we only have the derived azimuthal angle for each event (in practice there could be some possibility of limited position sensitivity along the anode wire).

The next step is to get the energy integrated modulation pattern, for which it is necessary to have knowledge of the incident spectrum. Here we carry out further analysis by assuming a power-law spectrum with the photon index ranging from -1.0 (very hard spectrum) to -3.0 (soft spectrum). The azimuthal distribution patterns obtained in the first step are normalized according to the expected source counts for the given type of source spectrum and given exposure time. The normalized patterns are then co-added to get the final energy integrated modulation pattern at given scatterer thickness. For the present analysis, we do not consider energy wise modulation pattern as a significant fraction of the photons are scattered by Compton scattering, which also gives similar modulation but alters the photon energy. The energy integrated modulation pattern provides simulated modulation factor for the 100% polarized photons for each scattering geometry. With this modulation factor and the overall detection efficiency (obtained from total number of detected photons), the MDP can be calculated for every geometry combination as described in section 2.2. However, first we calculate Figure-of-Merit (FoM), defined as $-\ FoM = \mu_{100}\sqrt{\varepsilon}$, for each geometry and scatterer thickness. The FoM as defined here is inversely proportional to the MDP but does not depend on the external factors such as source intensity, instrument background, exposure time etc. Thus, it is very useful for the direct comparison across various geometry combinations. Then we calculate MDP values for a range of source intensities and exposure time. The exact detector background ($R_{bkg}$) required for the MDP calculation is not known. However, form our earlier experience of space experiments with proportional counters [15,16], we expect it to be within the range of 0.01–0.1 counts cm$^{-2}$ s$^{-1}$. We consider both these extreme values for calculating MDP.

## 5. Results:

Main objective of our simulations is to determine $\mu_{100}$ and detection efficiency for various scattering geometries, scatterer material as well as scatterer thickness. Figure 3 (right panels) shows the modulation patterns for each geometry. Table 2 gives the actual modulation factor and detection efficiencies for all geometries and scattering materials for typical thicknesses. Figure 4 shows the variation of the figure-of-merit with scatterer thickness. It can be seen that there are significant differences between Be scatterer and Li / LiH scatterer, however Li and LiH scatterer show almost same behavior with LiH being slightly better. Also the dependence on scatterer thickness is rather weak. In order to compare sensitivities in terms of MDP (where additional variables of background rate, exposure time and spectral type are introduced) we select the scatterer thickness of 3 cm for Be, 8 cm for Li and 10 cm for LiH. Figure 5 shows the polarization sensitivity (in terms of 3 sigma MDP) for Crab like spectra (power-law index of -2.1), with 1 Ms (solid lines) and 100 ks (dashed lines) exposure times where as Figure 6 shows the sensitivity as a function of the source spectral index for 100 mCrab source with background rates of 0.01 counts s$^{-1}$ cm$^{-2}$ (solid lines) and 0.1 counts s$^{-1}$ cm$^{-2}$ (dashed lines). It can be seen that there are significant variations in various scattering geometries as well as scattering materials. The three dimensional scatterer (CONE and PYRD) provide better sensitivities than the two dimensional (DISC and SLAB) scatterer. However, there is not much difference between two shapes in each case. Further the cylindrical detectors provide significantly better sensitivity compared to square detector and thus the CYLL-CONE being the best combination. It can be seen that with Li or LiH scatterer, CYLL-CONE combination can achieve the desired sensitivity of 2% MDP for 100 mCrab source in 1 Ms exposure even for the higher end of the internal background rate. However with the Be scatterer, the actual detector background becomes the critical factor in determining the real sensitivity. The next best geometry combination is 4PPC-CONE, though there is not much difference between 4PPC-CONE and 4PPC-PYRD. Based on these results, the order of preference for the scattering geometries would be – CYLL-CONE, 4PPC-CONE, 4PPC-PYRD, CYLL-DISC, 4PPC-DISC and 4PPC-SLAB. However, the difficulty in realization for both the detector and scatterer increases in reverse order. Thus the final selection of the scattering geometry will have to be based on more practical aspects. The dashed lines in Figure 5 represent 100 ks exposure which shows that for bright sources the desired MDP can be achieved in much shorter time scale.

## 6. Summary and future work:

We have proposed an X-ray polarization measurement experiment onboard Indian Space Research Organization's (ISRO) small satellite platform. This experiment will be based on the principle of X-ray polarization measurement by Thomson scattering and will use Xenon filled proportional counters as the primary X-ray detectors. The performance of a scattering based X-ray polarimeter significantly depends on the scattering geometry. There are number of different experiment configurations possible for the proposed experiment which can be considered within the overall constraints of the small satellite platform. In order to take an informed decision about the final scattering geometry for the experiment, it is essential to have a comparative study of the performance of the various possible geometries. Here we have presented the results of our extensive Monte Carlo simulations carried out for this purpose. We find that various configurations considered here show significant variations in their sensitivity. The CYLL-CONE configuration with Li or LiH scatterer can achieve the proposed goal of 2% MDP for a 100 mCrab source in 1 million second exposure. However, in cases of other configurations, the desired sensitivity can be achieved only if the low detector background can be realized. It should be noted that the final experiment configuration will heavily depend on practical implementation feasibility of the particular configuration. However, these results will provide crucial inputs for taking an informed decision in this regard. Once the actual experiment configuration is finalized and more details are available about the spacecraft bus itself, we plan to extend the present work into a full fledge mass model of the experiment. This will be very useful to have a realistic estimate of the in-orbit background for the detector, which in turn will provide more accurate sensitivity estimates for the experiment.

**Table 1:** Exact dimensions used to construct scattering geometries for the present simulations.

| Proportional Counters | | | | |
|---|---|---|---|---|
| Square detector | Physical size | Width: 40 cm | Height: 40 cm | Thickness: 5 cm |
| | Active volume | Width: 30 cm | Height: 30 cm | Thickness: 3 cm |
| | Window support grid | Width: 32 cm | Height: 32 cm | Thickness: 1.2 cm |
| | Grid cells | Width: 2.8 cm | Height: 2.8 cm | Rib thickness: 0.2 cm |
| Cylindrical detector | Physical size | Outer radius: 25 cm | Inner radius: 20 cm | Height: 32 cm |
| | Active volume | Outer radius: 23 cm | Inner radius: 20 cm | Height: 30 cm |
| | Window support grid | Outer radius: 20 cm | Thickness: 1.2 cm | Height: 32 cm |
| | Grid cells | Height: 5.5 cm | Ang. width: 14.5 deg | Rib thickness: 0.3 cm |

| Scatterer | | | |
|---|---|---|---|
| Disc | | Radius: 18.0 cm | |
| Slab | | Size: 36.0 x 36.0 cm$^2$ | |
| Cone (hollow)[*] | | Outer bottom radius: 18.0 cm | Inner top radius: 2.0 cm |
| Pyramid (hollow)[*] | | Outer bottom square: 36 x 36 cm$^2$ | Top inner square: 2 x 2 cm$^2$ |

* In case of Cone and Pyramid scatterer inner bottom and outer top dimensions are decided by the scatterer thickness which is a variable for our simulations.

**Table 2:** Detection efficiency (Eff) and modulation factor (ModF) of scattered photons obtained from our simulations for various scattering geometries and scattering materials.

| | BE | | LI | | LIH | |
|---|---|---|---|---|---|---|
| | Eff (%) | ModF | Eff(%) | ModF | Eff(%) | ModF |
| CYLL-CONE | 2.02 | 40.28 | 3.59 | 41.00 | 4.71 | 35.46 |
| CYLL-DISC | 1.35 | 27.68 | 2.98 | 30.32 | 3.93 | 26.59 |
| 4PPC-CONE | 1.12 | 45.95 | 2.00 | 47.50 | 2.62 | 41.60 |
| 4PPC-PYRD | 1.06 | 39.51 | 1.88 | 41.00 | 2.46 | 35.20 |
| 4PPC-DISC | 0.70 | 34.52 | 1.57 | 37.90 | 2.09 | 32.98 |
| 4PPC-SLAB | 0.68 | 30.98 | 1.51 | 32.36 | 2.01 | 27.33 |

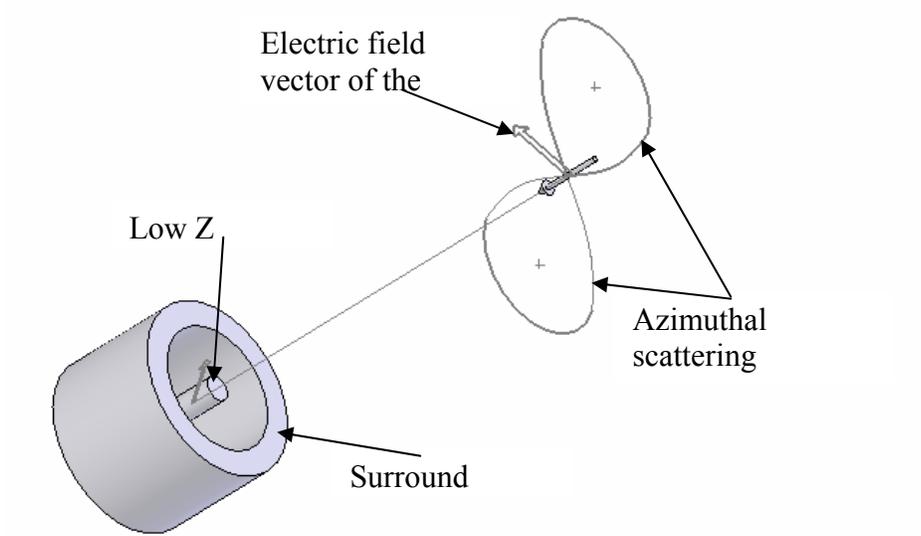

**Figure 1** Working principle of Thomson / Compton scattering based polarimeter. The azimuthal direction of the scattered photon depends on the polarization direction of the incident radiation. Thus by measuring the azimuthal scattering pattern with a detector surround scatterer, polarization of the incident radiation can be measured.

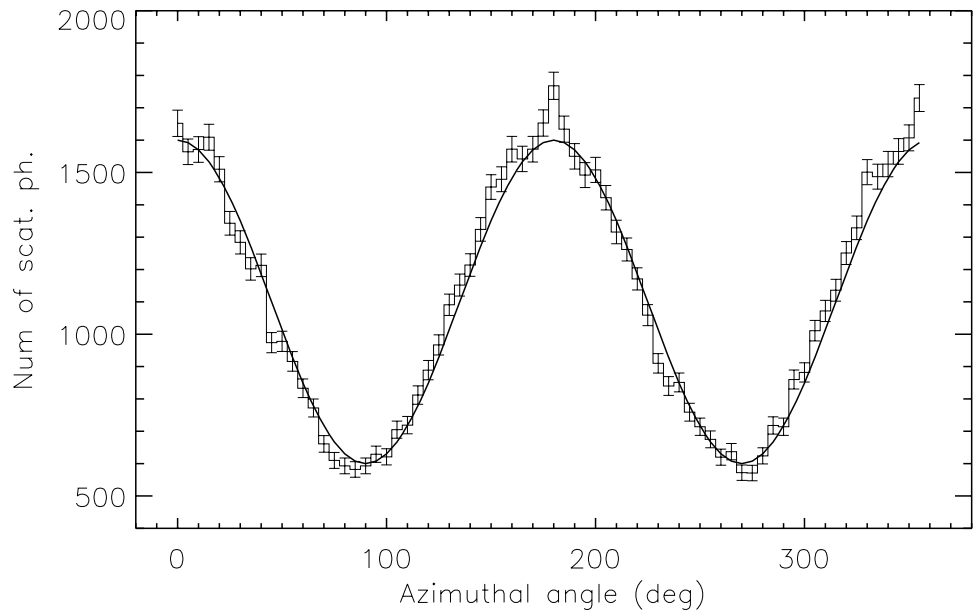

**Figure 2** Verification of the low energy polarized processes implemented in Geant4. The histogram represents simulated azimuthal modulation of the scattered photons from an ideal scatterer (long and narrow tube). The red line represents expected cos(2φ) modulation.

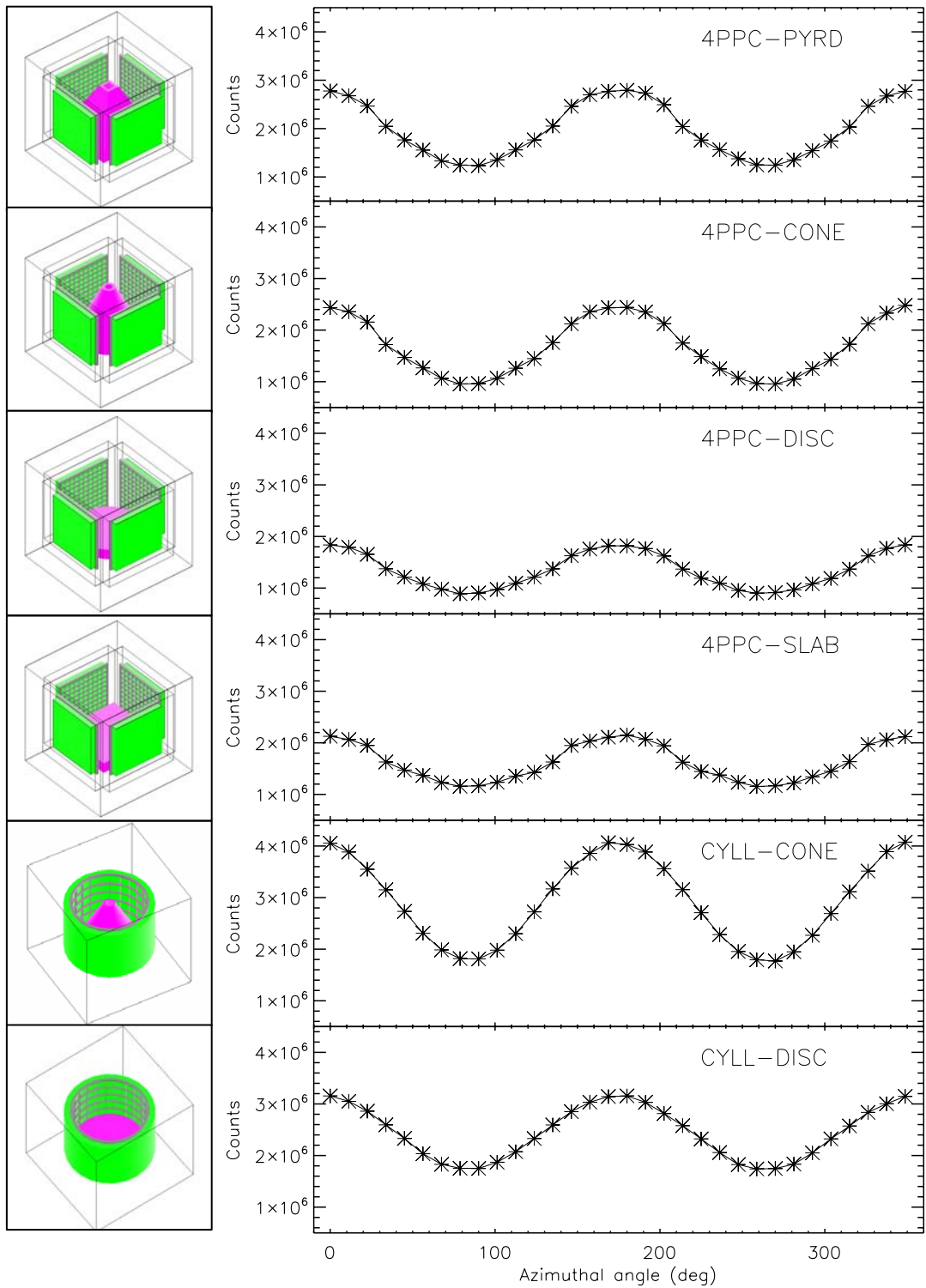

**Figure 3** Six scattering geometries as constructed in Geant4 and corresponding simulated modulation pattern.

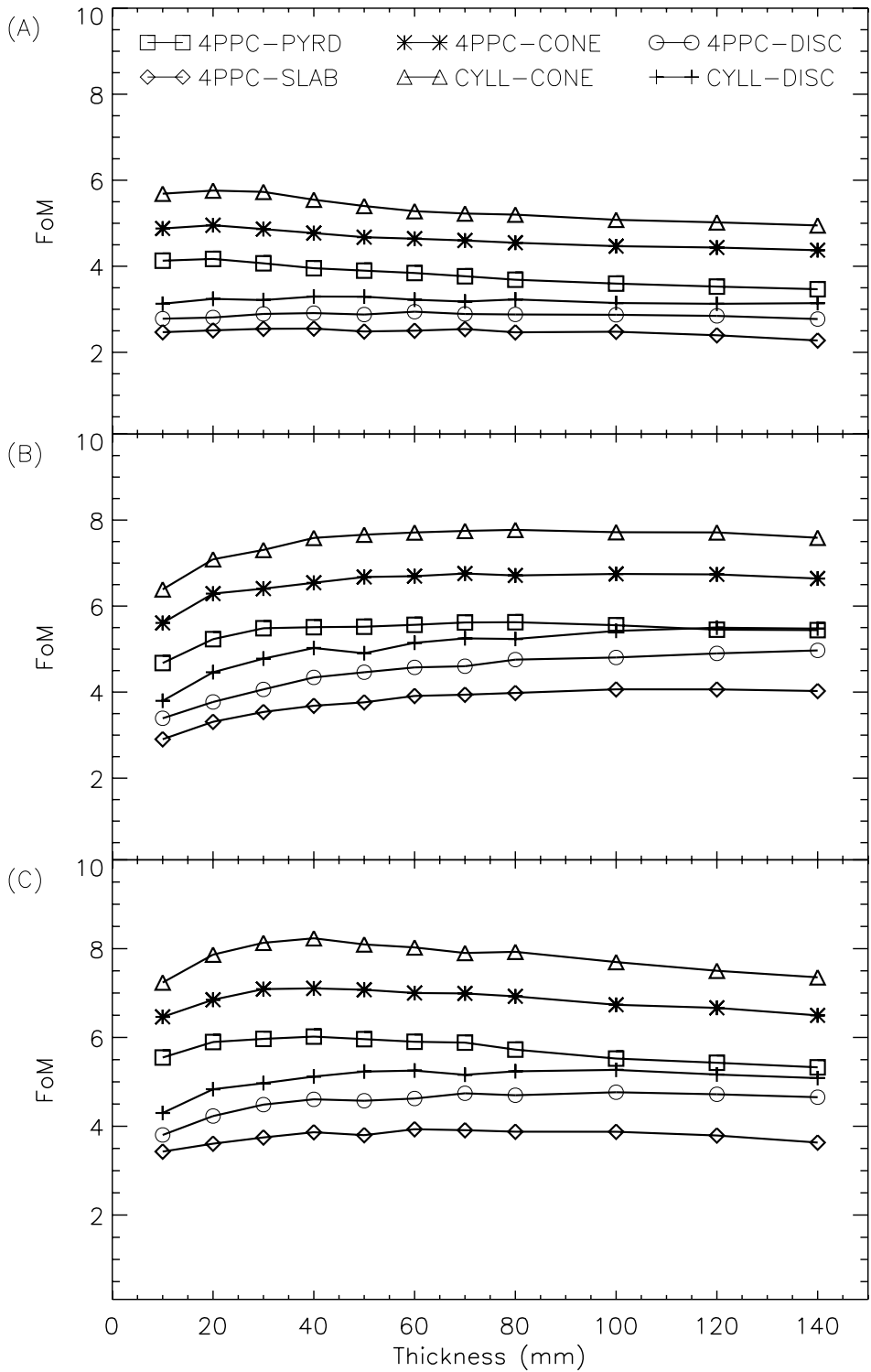

**Figure 4** Figure-of-Merit as a function of scatterer thickness for all six scattering geometries for (a) Be scatterer, (b) Li scatterer and (c) LiH scatterer. Symbols in all three panels have same meaning.

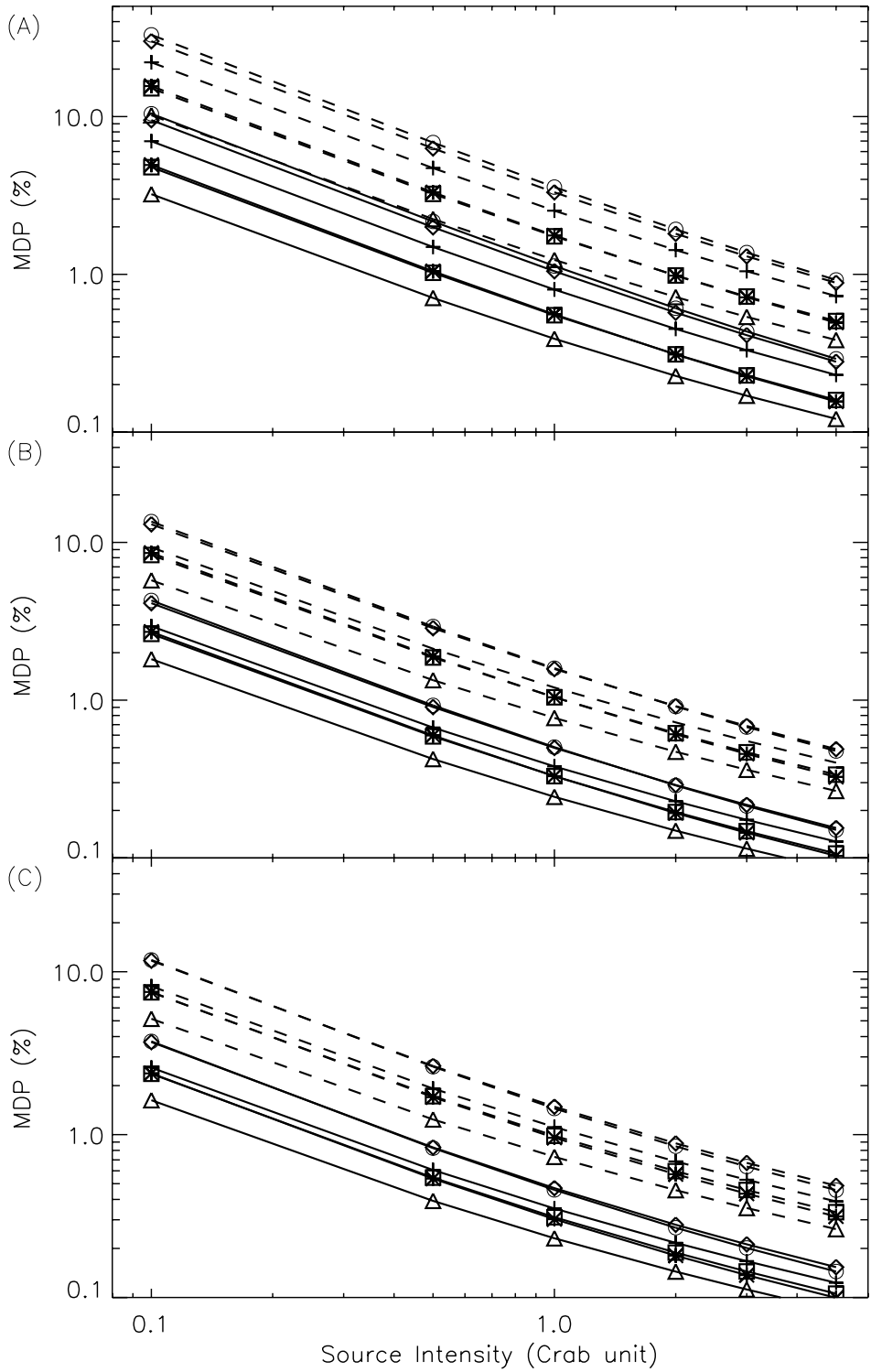

**Figure 5** Minimum Detectable Polarization (MDP) as a function of source intensity for all six scattering geometries with (a) Be scatterer (b) Li scatterer and (c) LiH scatterer. The solid lines represent 1 Ms exposure and dashed lines represent 100 ks exposure. The detector background rate is assumed to be 0.01 counts s$^{-1}$ cm$^{-2}$. Symbols have same meaning as Figure 4.

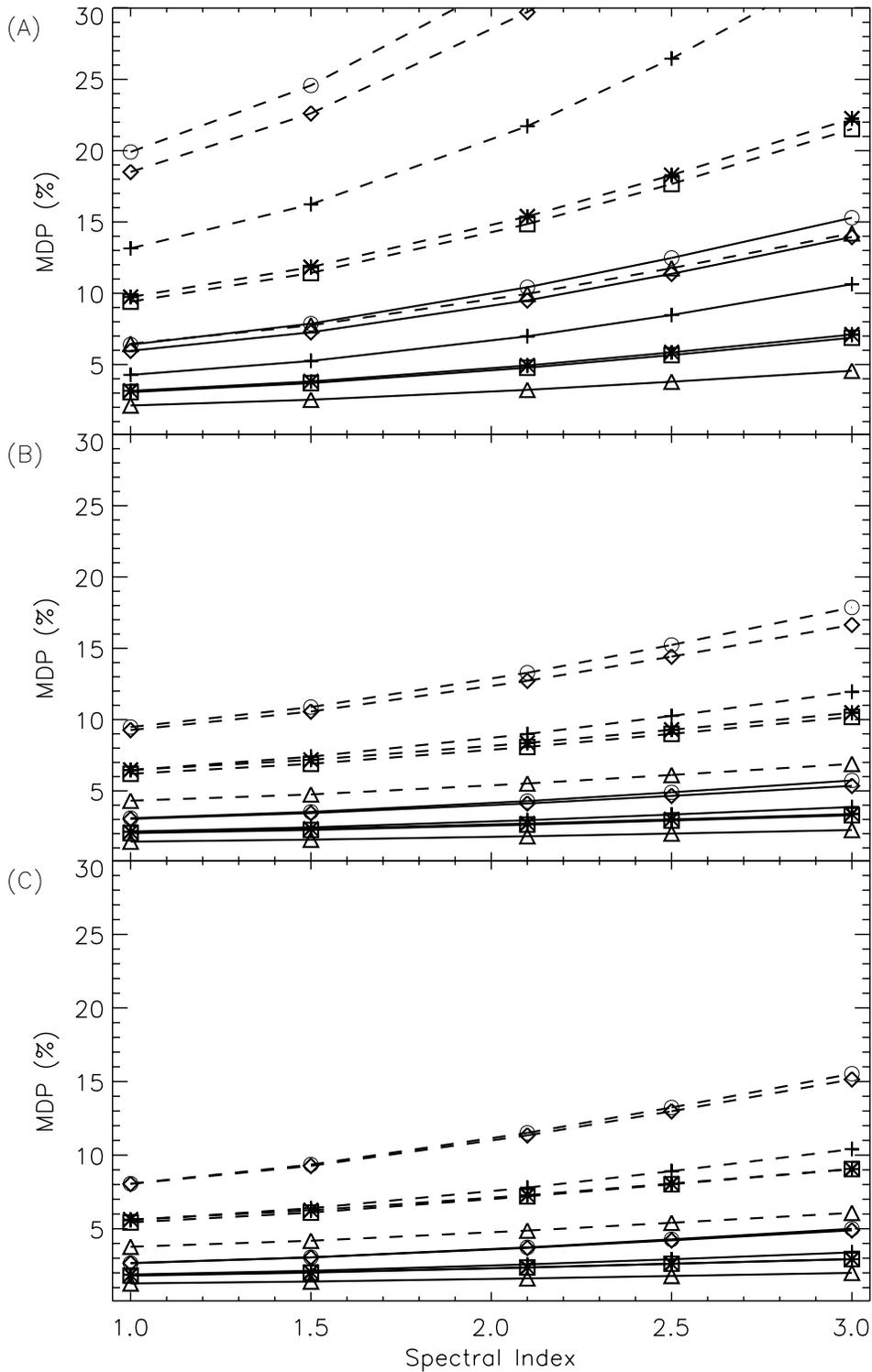

**Figure 6** Minimum Detectable Polarization as a function of source spectral index with source intensity of 100 mCrab and exposure of 1 Ms for (a) Be scatterer, (b) Li Scatterer And (c) LiH scatterer. The solid lines represent low background rate of 0.01 counts s$^{-1}$ cm$^{-2}$ whereas the dashed lines represent high background rate of 0.1 counts s$^{-1}$ cm$^{-2}$. Symbols have same meaning as Figure 4.